\documentclass{iau} 
\usepackage[rightcaption]{sidecap}
\usepackage{graphicx}

\title[Exotic AGN behind the Magellanic Clouds] 
{Discovering Exotic AGN behind the Magellanic Clouds}

\author[Clara M. Pennock \& Jacco Th. van Loon  ]   
{Clara M. Pennock$^1$, Jacco Th. van Loon$^1$, Cameron Bell$^2$, Miroslav Filipovi\'c$^3$,  Tana Joseph$^4$, \and Eleni Vardoulaki$^5$ }

\affiliation{$^1$Lennard-Jones Laboratories, Keele University, Keele, ST5 5BG, UK \\ email: {\tt c.m.pennock@keele.ac.uk} \\[\affilskip]
$^2$Leibniz Institute for Astrophysics Potsdam, Potsdam, Germany;
$^3$Western Sydney University, Sydney, Australia;
$^4$University of Manchester, Manchester, UK;
$^5$Max-Planck-Institut f{\"u}r Radioastronomie, Bonn, Germany.}

\pubyear{2019}
\volume{356}  
\jname{Nuclear Activity in Galaxies Across Cosmic Time}
\editors{M. Povi\'c, P. Marziani, J. Masegosa, H. Netzer,\\ S. H. Negu, \&
	S. B. Tessema, eds.}
\begin{document}

\maketitle

\begin{abstract}
The nearby Magellanic Clouds system covers more than 200 square degrees on the sky. Much of it has been mapped across the electromagnetic spectrum at high angular resolution
and sensitivity –X-ray (XMM-Newton), UV (UVIT), optical (SMASH), IR (VISTA, WISE, Spitzer, Herschel), radio (ATCA, ASKAP, MeerKAT). This provides us with an excellent
dataset to explore the galaxy populations behind the stellar-rich Magellanic Clouds. We seek to identify and characterise AGN via machine learning algorithms on this
exquisite data set. Our project focuses not on establishing sequences and distributions of common types of galaxies and active galactic nuclei (AGN), but seeks to identify
extreme examples, building on the recent accidental discoveries of unique AGN behind the Magellanic Clouds.
\end{abstract}

\firstsection 
\vspace{0.1mm}
\section{Newest Data}
The Vista survey of the Magellanic Clouds (VMC; \cite{cioni2011}) is a recently completed deep-field multi-epoch, near-IR survey taken with the 4m VISual and Infrared Telescope for Astronomy (VISTA) at the European Southern Observatory at Cerro Paranal, Chile. It observed in the $Y$, $J$ and $K_s$ bands, going to a depth of $K_s$=22.2 mag (AB). It covers the Large Magellanic Cloud (LMC) area ($116$ ${\rm deg}^2$), the Small Magellanic Cloud (SMC) area ($45$ ${\rm deg}^2$), the Bridge ($20$ ${\rm deg}^2$) and two tiles in the Stream ($3$ ${\rm deg}^2$).

The Evolutionary Map of the Universe (EMU; \cite{norris2011}) is a wide-field radio continuum survey which uses the Australian Square Kilometre Array Pathfinder (ASKAP). It has a frequency range from 700 MHz to 1.8 GHz and is capable of a $30$ ${\rm deg}^2$ field-of-view. EMU's primary goal is to make a deep (rms $\sim$ 10 $\mu$Jy/beam) radio continuum survey of the Southern sky, with 10 arcsec resolution. The SMC survey (\cite{joseph2019})) was taken as part of ASKAP early science verification. The LMC was also observed and is currently in the data processing stage. 

The combination of these two wavelength ranges will provide a powerful tool for searching for AGN, in particular towards higher redshifts and/or obscured AGN.
\vspace{-4mm}
\section{The Exotic}
SAGE1C J053634.78$-$722658.5 (SAGE0536AGN) is an abnormal AGN discovered behind the LMC in the Spitzer Space Telescope “Surveying the Agents of Galaxy Evolution” Spectroscopy survey (\cite{hony2011}) and has since been better characterised with the Southern African Large Telescope (SALT) (\cite{vanloonsansom2015}). Sitting at redshift $z=0.1428 \pm 0.0001$, the striking feature of this AGN is its uniquely strong silicate emission at 10 $\mu$m and its lack of star-formation activity. From our use of Galfit (\cite{galfit}) it has been found that this is a disc galaxy instead of an early-type galaxy as previously thought.

Another interesting source is a striking X-shaped radio AGN candidate (\cite{joseph2019}). Predicted to be a binary AGN, seen just before the AGN merge and consisting of two jets emitting in parallel and opposite directions from the central source. The third and middle jet is relativistically boosted towards us.
We also found a near-IR look-a-like using WISE colour-colour diagrams cross-matched with EMU radio sources and VMC images. These sources need to be disentangled and we are using the combination of the VMC and optical spectroscopy from SALT to ascertain the true natures of each of the radio counterparts. See Figure 1 for these two radio sources.
\begin{figure*}[h!]
\centering
\includegraphics[scale=0.5]{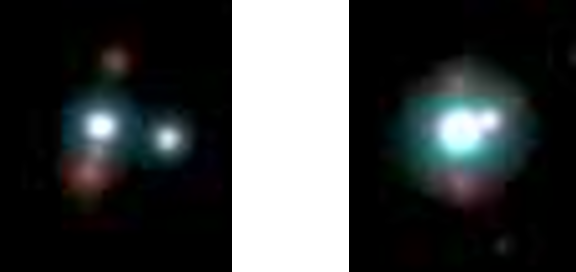}
\caption{(Left) Original X-shaped radio AGN candidate and (right) near-IR look-a-like. Colours represent $Y$ (blue), $J$ (green), $K_s$ (red) from the VMC. }
\label{fig:RF}
\end{figure*}

\vspace{-4mm}
\section{Random Forest Basics}
Supervised machine learning algorithms predict classifications/values based off example data with known classifications/values. It does this by analysing a known sample, the training set, and producing a model, which can then be used to make predictions of the output of an unseen dataset. 

 Decision trees (\cite{breiman2001}) predict the value/class of a target variable by creating a model that has learnt simple decision rules inferred directly from the data features it is trained upon. It consists of nodes where a condition is given that is either true or false. The answer to this condition leads down a “branch” to the next node and condition, where either another split happens or the output variable is given.

An ensemble of decision trees is called the random forest algorithm (\cite{breiman2001}), which  can be used for both classification and regression problems. The algorithm builds several decision trees independently and then averages the predictions of these to obtain the final prediction. This reduces variance over using a single estimator and creates an overall more stable model. Randomness is injected into the training process of each individual tree via a method called 'bagging'. This method splits up the training set into randomly selected subsets, and each decision tree is then trained on one of those subsets. 

Advantages of the random forest include: they require very little data preparation; it constructs a non-linear model during training, which is advantageous as most problems require non-linear solutions. It can handle numerous features and objects. It can produce probabilities of each class for each prediction (fuzzy logic) and produce feature importance, showcasing which features have more bearing on the classifications it had made. Lastly, it generalises well to unseen datasets due to the inclusion of randomness.
\vspace{-4mm}
\section{The Training and Results}
The features used in the random forest are magnitudes/fluxes and colours from survey catalogues: EMU (\cite{norris2011}), unWISE (\cite{schlafly2019}), SAGE (LMC --- \cite{meixner2006}, SMC/Bridge --- \cite{gordon2011}), VMC (Cioni et al. 2011), Gaia DR2 (Brown et al. 2018), GALEX (\cite{bianchi2017}), XMM-Newton (\cite{sturm2013}) and including features calculated from VMC data such as source density and sharpness. It is trained on 75$\%$ of the training
data and tested on remaining 25$\%$.

The output classifications are: AGN (spectroscopically confirmed), Galaxies (no AGN, spectroscopically confirmed), Stars -– Proper Motion/Emission/RGB/Carbon/Red Supergiant (SG), Young Stellar Object (YSO).

The resulting trained random forest produced
AGN accuracies ranging from 80 -- 90$\%$. While AGN can be
misclassified as other sources, the predicted AGN tend
to be correct, with a false positive rate of 1 -- 2$\%$. The current limitations are caused by limited coverage of X-ray and UV catalogues that avoid the central parts of the Magellanic Clouds. Furthermore, only 26.7$\%$ and 7.5$\%$ of the known AGN sources are X-ray and radio sources, respectively. As these emission types are often indications of an AGN, their lack of representation in the training set lessens the likelihood the random forest will see these emissions as clear indications of AGN.  
\vspace{-4mm}
\section{Future Work}
Future investigations will look into the difference between the correctly classified and the misclassified AGN.
Optical spectroscopic observations using SAAO’s (South
African Astronomical Observatory) SALT and 1.9m telescopes to increase the number of AGN in the training set from behind the
 Magellanic Clouds are currently underway. This is, however, biased towards the more unobscured sources.
We have proposed to use the ESO NTT to observe
spectroscopically in the infrared the
higher redshift sources that, while many of the radio sources are detected in the VMC, their optical counterparts are all invisible except for nearby, unobscured and/or extreme cases.
We will also incorporate optical SMASH data that has just become
publicly available into the machine learning as an
improvement over Gaia data and augment the UV GALEX
features with UVIT data. In addition, we will include variability data from
VMC into machine learning as well as features based on
the environment around the sources and account for
reddening (by dust) and redshift.
Lastly, we will employ the use of a probabilistic random forest, which allows
the addition of uncertainty in the features and labels of
the sources and missing data. It has been shown to
provide greater accuracy in predictions (\cite{reis2018}). Current tests of using it, however, have yielded
lower accuracy's in the prediction of AGN and this needs to
be improved upon.


\vspace{-4mm}
\begin{thebibliography}{}
\bibitem[Norris \etal\ 2011]{norris2011}
{Norris, R.~P., Hopkins, A.~M., Afonso, J., et al.,} 2011, 
\textit{PASA}, 28, 215

\bibitem[Joseph \etal\ 2019]{joseph2019}
{Joseph, T.~D., Filipovi{\'c}, M.~D., et al.,} 2019, 
\textit{MNRAS}, 490, 1202

\bibitem[Hony et al. 2011]{hony2011}
{Hony, S., Kemper, F., Woods, P.~M., et al.,} 2011, 
\textit{AAP}, 531, A137

\bibitem[van Loon \& Sansom, 2015]{vanloonsansom2015}
{ van Loon, J.~T., \& Sansom, A.~E.,} 2015, 
\textit{MNRAS}, 453, 2341

\bibitem[Breiman 2001]{breiman2001}
{Breiman, L.,} 2001, 
\textit{Machine Learning}, 45, 1

\bibitem[Schlafly et al. 2019]{schlafly2019}
{Schlafly, E.~F., Meisner, A.~M., \& Green, G.~M.,} 2019, 
\textit{APJS}, 240, 30

\bibitem[Meixner et al. 2006]{meixner2006}
{Meixner, M., Gordon, K.~D., Indebetouw, R., et al.,} 2006, 
\textit{AJ}, 132, 2268

\bibitem[Gordon et al. 2011]{gordon2011}
{Gordon, K.~D., Meixner, M., Meade, M.~R., et al.,} 2011, 
\textit{AJ}, 142, 102

\bibitem[Cioni et al. 2011]{cioni2011}
{Cioni, M.-R.~L., Clementini, G., Girardi, L., et al.,} 2011, 
\textit{AAP}, 527, A116

\bibitem[Gaia Collaboration et al.2018]{gaia2018}
{ Gaia Collaboration, Brown, A.~G.~A., Vallenari, A., et al.,} 2018, 
\textit{AAP}, 616, A1

\bibitem[Bianchi et al. 2017]{bianchi2017}
{Bianchi, L., Shiao, B., \& Thilker, D.,} 2017, 
\textit{APJS}, 230, 24

\bibitem[Sturm et al. 2013]{sturm2013}
{Sturm, R., Haberl, F., Pietsch, W., et al.,} 2013, 
\textit{AAP}, 558, A3

\bibitem[Reis et al. 2018]{reis2018}
{Reis, I., \& Baron, D. \& Shahaf, S.,} 2018, 
\textit{AJ}, 157, 16 

\bibitem[Peng et al. 2002]{galfit}
{Peng, C.Y., Ho, L.C., Impey, C.D., \& Rix, H.-W.,} 2002,
\textit{AJ}, 124, 266

\end{thebibliography}
\end{document}